\documentstyle[multicol,epsf,aps,pre]{revtex}
\begin{document}
\title{Infiltration through porous media}
\author{W. Hwang and S. Redner}
\address{Center for BioDynamics, Center for Polymer Studies, 
and Department of Physics, Boston University, Boston, MA 02215}

\maketitle 
\begin{abstract} 
  We study the kinetics of {\em infiltration} in which contaminant particles,
  which are suspended in a flowing carrier fluid, penetrate a porous medium.
  The progress of the ``invader'' particles is impeded by their trapping on
  active ``defender'' sites which are on the surfaces of the medium.  As the
  defenders are used up, the invader penetrates further and ultimately breaks
  through.  We study this process in the regime where the particles are much
  smaller than the pores so that the permeability change due to trapping is
  negligible.  We develop a family of microscopic models of increasing
  realism to determine the propagation velocity of the invasion front, as
  well as the shapes of the invader and defender profiles.  The predictions
  of our model agree qualitatively with experimental results on breakthrough
  times and the time dependence of the invader concentration at the output.
  Our results also provide practical guidelines for improving the design of
  deep bed filters in which infiltration is the primary separation mechanism.

\end{abstract}
\begin{multicols}{2}
\narrowtext
\section{Introduction}
In depth filtration, suspended particles in a fluid are removed during their
passage through a porous medium\cite{revs,probstein}.  The basic dynamics of
depth filtration is determined primarily by the pore structure of the filter,
the particle size distribution, and by various physicochemical and
hydrodynamic details.  If the particle size is larger than the typical pore
size, particles get stuck relatively quickly.  The permeability of the filter
decreases steadily during this process and drops to zero when clogging is
reached.  This process is often referred to as sieving, or
straining\cite{sahimi}.  Conversely, if particles are much smaller than the
pore size and if particles are trapped only at the interfaces of the porous
medium, the flow field is only slightly affected by the trapping.  The goal
of this paper is to provide a general understanding of this latter process of
{\em infiltration} by microscopic network modeling.

\begin{figure}
\narrowtext
\hskip2.0cm\epsfxsize=35mm\epsfbox{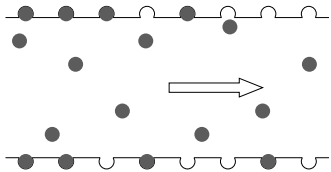} \vskip.4cm\
\caption{Idealized picture of infiltration.  Suspended particles are
  trapped at ``defender'' sites on pore surfaces.  Once defenders are
  occupied, subsequent particles pass by freely.}
\label{fig0}
\end{figure}

Infiltration underlies many practical situations, such as underground waste
disposal \cite{luhrmann}, gas mask design, or drinking water filters
\cite{bloomfield}.  Typically, sub-micron size contaminant particles are
suspended in a carrier fluid and flow through a porous material, such as a
sand filter whose typical grain size is much larger than the contaminant
particles, or an ion exchange filter\cite{bloomfield} where the contaminant
size is molecular in scale.  In such cases, one can neglect the change of the
flow field due to particle trapping \cite{luhrmann}, an approximation which
considerably simplifies theoretical analysis.

The kinetics of infiltration is controlled by the microscopic mechanisms for
the trapping of the invader particles.  Typically each pore can hold a
limited number of particles due to a finite surface area or a finite range of
the surface potential.  When all the available surface area is covered by
particles, subsequent invaders flow passively through the filter without
being trapped.  Our basic goal is to understand the kinetics of this
infiltration and the ultimate breakthrough of the invader, as well as the
evolution of the invader and defender density profiles as functions of
downstream position and time.

Previous work on infiltration in porous media has often been based on a
macroscopic convection-diffusion equation description, with reaction terms
introduced to account for particle trapping \cite{luhrmann,putnam,brenner}.
Another approach has been to use a single absorbing sphere to calculate the
collection efficiency at the initial stage of filtration\cite{rajagopalan}.
While numerical simulations of these models have some predictive power, it is
hard to develop a connection between this macroscopic approach and basic
features of the microscopic process, such as the concentration profiles of
the trapped and flowing particles.

For filtration by straining, models based on a discrete network description
of the filter medium are relatively well developed
\cite{jysoo,ghidaglia,datta,imdakm,rege1}.  To our knowledge, however, there
has been no microscopic network modeling work on infiltration.  As in the
case of straining, a spatial density gradient naturally arises in
infiltration, since particles begin to deposit at the upstream end of the
filter and advance downstream as the filter gets used up. The density
gradient is experimentally observed as the time-dependent output
concentration \cite{putnam}. In this paper, we will account for this basic
experimental observation by using a discrete network approach.

Practical questions raised by infiltration are the breakthrough time, which 
is defined as the time for the output concentration to reach a specified 
threshold level, and the filter efficiency, which is related to the fraction 
of the filter material actually used before breakthrough.  Clearly, it is 
desirable to use as much of the filter material as possible before 
breakthrough occurs. 

This paper is organized as follows.  In Sec.~\ref{basic}, we introduce the
basic parameters that govern particle trapping and provide a qualitative
picture of infiltration.  In the following sections, we construct a sequence
of discrete models with increasing complexity and realism to ultimately
provide a lattice network description.  In Sec.~\ref{1d}, we discuss the case
of a one-dimensional (1D) chain of trapping sites and in Sec.~\ref{bubble},
we analyze infiltration in the bubble model to provide a mean-field-like
description.  Building on these results, we then turn to simulations of
infiltration on tube lattice networks in Sec.~\ref{network}.  We summarize
and compare our results with experiments in Sec.~\ref{discussion}.


\section{Basic Picture}\label{basic}

\begin{figure}
\narrowtext
\hskip0.25cm\epsfxsize=60mm\epsfbox{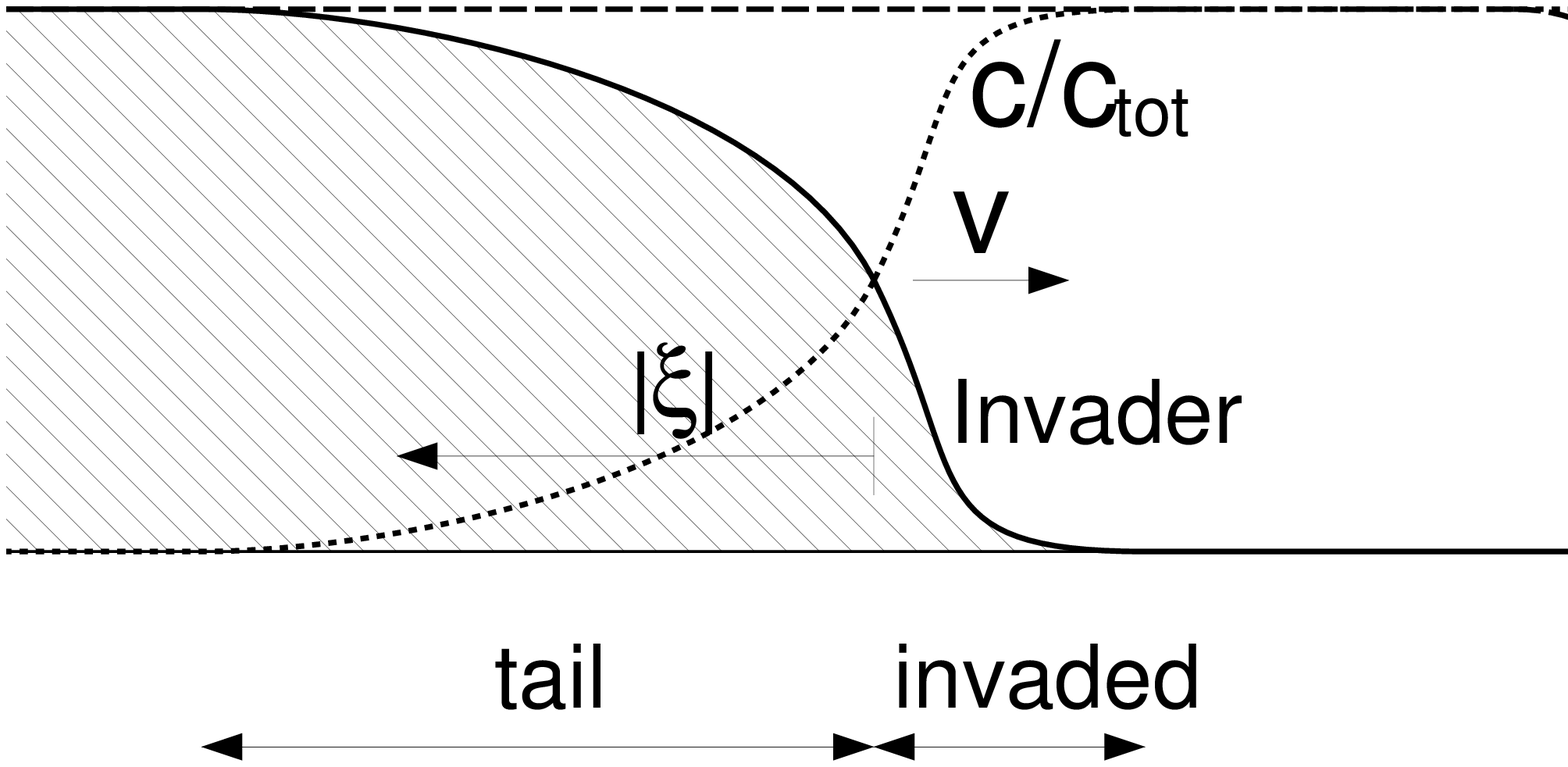} \vskip.6cm 
\caption{Infiltration profiles.  Horizontal direction is downstream. 
  The invader density profile (shaded) is exponential in the invaded region,
  while the capacity profile has a power law tail, $c\sim|\xi|^{-\lambda}$.}
\label{fig1}
\end{figure}

The two basic characteristics of particle trapping are the efficiency of an
unoccupied trapping site and the number of trapping sites in a pore.  We
introduce the trapping probability $\gamma$ as the probability that a
particle is trapped upon encountering an open collector site.  The parameter
$\gamma$ thus represents the strength of the particle-collector interaction
and accounts for the possibility that contact between particles and the
filter grains may not necessarily lead to deposition \cite{chiang}.  While
this simplifies the complicated adsorption mechanism, later we show that the
basic feature such as the invasion front propagation velocity
(Fig.\ref{fig1}) is independent of the interaction details.

Next we introduce the capacity $c_x(t)$ as the number of particles a pore at
position $x$ can hold at time $t$.  In the case of non-coagulating particles
which cannot get trapped on top of an already adsorbed particle, the initial
capacity is proportional to the inner surface area of the pore and then
decreases as the pore surface is covered by particles.  For simplicity, we
ignore multiple trapping on an already occupied collector site as well as
particle re-launching.  The key factors which determine the dynamic behavior
of the system are geometric, such as the capacity of a clean filter and the
pore size distribution, and kinematic, such as the particle concentration and
the flow rate.  More refined models for particle trapping can be incorporated
within our basic modeling.

Consider a generic infiltration process based on the above concepts.
Initially a layer in the clean filter has a total capacity $c_{\rm tot}$
independent of the downstream position. At $t=0$, a fluid which contains a
mixture of invader and non-reacting tracer particles enters the filter whose
flow rate is determined by the steady-state solution of d'Arcy's law. Tracer
particles passively follow the fluid motion and advance with the average flow
velocity $v_0$.  The width of the tracer density profile spreads as $t^{1/2}$
due to hydrodynamic dispersion (Fig.~\ref{fig1}).  Invader particles first
encounter clean collector sites.  Because each such encounter leads to
deposition with probability $\gamma$, the survival probability of the
particles in this leading invaded region decreases exponentially with
downstream position as illustrated in Fig.~\ref{fig1}.  As particles advance
and get trapped, the pore capacity decreases and subsequent particles are
more likely to survive, giving rise to an advancing invasion front with a
velocity $v<v_0$.  In principle, the propagation velocity and shape of the
front are functions of time.  However, at long times these features approach
steady-state values.  Another important feature is that the trailing edge of
the capacity profile decays as a power of the distance $|\xi|$ 
($\xi<0$) from the
invasion front whose location is defined, {\it e.g.}, as the position 
where $c=c_{\rm tot}/2$.
For large $|\xi|$, any reasonable definition for the front location can be
used.
  
The existence of different propagation velocities, $v_0$ for the pure fluid
and $v<v_0$ for the contaminant, leads to purification of the liquid.  The
filter can be used until the invaded region reaches the outlet end.  For a
filter of length $L$, the breakthrough time will be of the order of $L/v$, so
the amount of throughput will be approximately proportional to $Lv_0/v$.

\section{One-dimensional model}\label{1d}

As a preliminary, we study infiltration in a one-dimensional chain of
identical pores at $k=0, 1, 2,\ldots$.  First we consider the case where each
pore can accommodate only one particle and then we generalize to multiple
capacity pores.

\subsection{Single capacity pores}\label{1dsingle}

We choose a time unit such that one particle is injected at each discrete
time step.  Multiple particle injection leads to a different particle density
and will affect only the overall scale factor and not change qualitative
features of the system.  The carrier fluid advances by one pore distance at
each time step; that is, its velocity is unity.  When particle trapping
occurs in a pore at $k$, the capacity $c_k(t)$ changes permanently from 1 to
0.  At time $t$, a particle at pore $k$ gets trapped in that pore with
probability $\gamma$ if $c_k(t)=1$.  A particle advances to the next pore in
one time step with probability 1 if $c_k(t)=0$.  Based on these elemental
steps, we introduce the following two probability densities:
\begin{itemize}
\item $p_k(t)$: The probability that a freely-moving particle is in pore $k$
  at time $t$.
\item $q_k(t)$: The probability that the pore at site $k$ is unoccupied; that
  is $c_k(t)=1$.
\end{itemize}
The corresponding master equations for $p_k(t)$ and $q_k(t)$ are
\begin{eqnarray} 
\label{1dmep}
p_k(t+1)&=&p_{k-1}\bigl( 1-\gamma q_{k-1}\,\bigr),\\
\label{1dmeq}
q_k(t+1)&=&q_k\bigl( 1-\gamma p_k\,\bigr),
\end{eqnarray} 
where we drop the argument $t$ on the right hand side for simplicity.  Unless
there is a possibility for confusion, we will not write the argument $t$ in
related formulae.  Since a particle advances to the next pore in one time
step, $p_k(t+1)$ depends on $p_{k-1}(t)$.  The term $(1-\gamma q_{k-1})$ in
Eq.~(\ref{1dmep}) is the probability that the particle at $k-1$ does not get
trapped by an unoccupied pore also at $k-1$.  Similarly, the term $(1-\gamma
p_k)$ in Eq.~(\ref{1dmeq}) is the probability that pore $k$ does not trap a
free particle at time $t$. The initial and boundary conditions for these
equations are:
\begin{equation} 
\label{1dic} 
p_{k\geq1}(0)=0\quad q_k(0)=1;\quad p_0(t)=1\quad q_\infty(t)=1
\end{equation} 

If the trapping probability $\gamma$ is small, a particle can advance many
pores without being trapped, so that $p_k(t)$ and $q_k(t)$ vary slowly in
space and time and a continuum approximation can be applied.  Letting $k+1\to
x+\delta x$ and $t+1\to t+\delta t$, Eqs.~(\ref{1dmep}) and (\ref{1dmeq})
become, to lowest order,
\begin{eqnarray}
\label{pq} 
\partial_t p+v_0 \partial_x p&=& -\gamma pq,\nonumber \\
\qquad\partial_t q   &=& -\gamma pq,
\end{eqnarray} 
where $v_0\equiv\delta x/\delta t=1$, and $\gamma\rightarrow\gamma/\delta t$
is a redefinition of the trapping probability in units of the infinitesimal
time increment.

In a co-moving reference frame, $\xi\equiv x-v t$, with $v$ the invasion
front propagation velocity shown in Fig.~\ref{fig1} (which is yet to be
determined), Eqs.~(\ref{pq}) become
\begin{eqnarray} 
\label{1dpwave}
(v_0 - v)\partial_\xi p+\partial_t p=-\gamma pq\\
\label{1dqwave}
-v\partial_\xi q+\partial_t q=-\gamma pq.
\end{eqnarray} 

Let us first examine the steady state solution of Eqs.~(\ref{1dpwave}) and
(\ref{1dqwave}).  Setting the time derivatives to zero, subtracting
Eq.~(\ref{1dqwave}) from Eq.~(\ref{1dpwave}), and integrating with respect to
$\xi$ gives
\begin{equation} 
\label{1dpqsol}
(v_0 -v)p(\xi)+vq(\xi)={\rm const}.
\end{equation} 
The integration constant can be determined by applying Eqs.~(\ref{1dic}) in
the co-moving frame.  As $\xi\to -\infty$, $p\to 1$, $q\to 0$, and as $\xi\to
\infty$, $p\to 0$, $q\to 1$.  These immediately give $v=v_0/2=0.5$.  Note
that $v$ is determined entirely from the boundary condition (and $v_0$) in the
co-moving frame, and not from the interaction strength $\gamma$.  This
feature continues to hold for all the models in this paper.

\begin{figure}
\narrowtext
\hskip8mm\epsfxsize=65mm\epsfysize=35mm\epsfbox{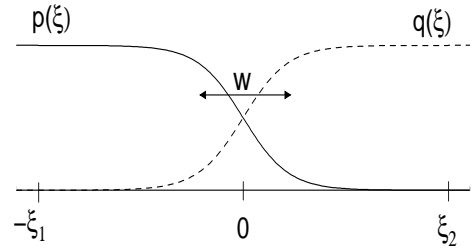} \vskip.1cm
\caption{Steady state profile in the single capacity model.} 
\label{fig2}
\end{figure}

Using Eq.~(\ref{1dpqsol}) with $v=v_0/2$, Eqs.~(\ref{1dpwave}) and
(\ref{1dqwave}) can be solved to give
\begin{equation}
\label{1dpq}
p(\xi)={1\over 1+e^{\xi/\xi_0}},\qquad q(\xi)={1\over 1+e^{-\xi/\xi_0}},
\end{equation}
(Fig.~\ref{fig2}).  Thus $\xi_0\equiv v_0/2\gamma$ is the characteristic 
width $w$
of the profile.  Notice also that the profiles of $p$ and $q$ are symmetric
about their intersection.  We verified both the dependence of the width on
$\gamma$ and the profile shape predicted by Eq.~(\ref{1dpq}) by numerical
integration of the master equations (\ref{1dmep}) and (\ref{1dmeq}).

One subtle point is the rate of approach to the steady state.  First, we find
that the asymptotic propagation velocity $v=v_0/2$ is reached before the
asymptotic profile is established.  This arises because $v$ is determined by
the boundary conditions, and not by interaction details.  We also verified
this feature numerically.  Adopted this asymptotic velocity,
Eqs.~(\ref{1dpwave}) and (\ref{1dqwave}) are then symmetric in $p$ and $q$.
In fact, the system is identical to two-species annihilation, $A+B\to 0$,
where each species is ballistically injected from opposite sides with
velocity $+v_0/2$ for the $A$s and $-v_0/2$ for the $B$s.  

Thus most of the the time variation in Fig.~\ref{fig2} occurs in the reactive
region of width $w$, where $w=0$ at $t=0$.  Integrating Eq.~(\ref{1dpwave})
from $-\xi_1\ll 0$ to $\xi_2\gg 0$ gives,
\begin{equation}
\label{1dpint}
-{v_0\over 2}+\partial_t \int_{-\xi_1}^{\xi_2} p\:d\xi
   =-\gamma\int_{-\xi_1}^{\xi_2} pq\:d\xi.
\end{equation}
The integral on the left hand side is the area under the curve $p(\xi)$
between $-\xi_1$ and $\xi_2$, whose time dependence mainly comes from the
change in $w$. On the right hand side, the integral is approximately
proportional to $w$, since $pq$ is significantly different from zero only in
the reactive region.  We can then rewrite Eq.~(\ref{1dpint}) as
\begin{equation}
\label{1dpint1}
-{v_0\over 2}+c_1\partial_t w \simeq -\gamma c_2 w,
\end{equation}
where $c_1$ and $c_2$ are constants.  Integrating Eq.~(\ref{1dpint1}) and
applying the condition $w(\infty)\sim {v_0\over2\gamma}$ gives an exponential
decay to the steady state $w(t)\sim {v_0\over2\gamma}(1 -e^{-c_2\gamma t})$.

It is worth emphasizing that the symmetry between the invader and defender is
generally responsible for the relation $v=v_0/2$.  At the inlet, invaders are
injected with velocity $v_0$, and the invasion front advances with velocity
$v$, with one invader particle annihilating with one defender site.  In the
reference frame moving with velocity $v_0$, the situation is reversed.  The
invaders are at rest and defenders are injected with velocity $v_0$ from the
opposite direction.  Therefore the invasion front advances with velocity
$v_0-v$.  Since these two reference frames describe the system in the same
way, the front velocities should be the same; that is, $v=v_0-v$, or
$v=v_0/2$.


\subsection{Multiple capacity pores}\label{1dmulti}

Now we consider the case where each pore can trap $M$ particles, that is, the
initial pore capacity is $c_k(0)=M$.  We again follow the previous rules of
injecting a single particle and advancing a particle by one pore ($v_0=1$) at
each time step.  Multiple particle injection or different injection intervals
again simply changes the overall concentration and time scale.  

In a multiple capacity pore, the probability of encountering an open trap in
a pore needs be considered, in addition to the trapping probability $\gamma$
upon encounter with an open trap. Generally, the encounter probability
decreases as more particles get trapped, since the inner pore surface area
available for trapping shrinks.  When fluid mixing within a pore is weak, a
particle can encounter only one trap, either open or occupied.  Then the
encounter probability is approximately proportional to the fraction of the
open surface area.  On the other hand, if the mixing is perfect, a particle
encounters all the available traps in a pore before exiting.

For practically relevant situations, pores are sufficiently short so that a
particle in a pore follows streamlines without transverse diffusive mixing
\cite{levich}.  In what follows, we consider this limit of weak mixing.  For
a pore with $n$ out of $M$ traps available, the encounter probability is
$n/M$, and the overall trapping probability of this pore is ${T}_n\equiv
\gamma n/M$.  In writing this expression, we ignore the possibility that a
particle far from the pore wall does not encounter any traps.  In
Sec.~\ref{discussion}, we argue that this volumetric effect does not change
the basic behavior of infiltration.

To describe the evolution of the system, we use the same single-particle
probability density $p_k(t)$ as in the single-capacity pore system, but
modify the probability density for the capacity as follows:
\begin{itemize}
\item $q_k^n(t)$: The probability that a pore at position $k$ contains $n$
  open traps. This is the same as the probability that $c_k(t)=n$, for $0\le
  n\le M$.
\end{itemize}
Following similar reasoning as that applied to deduce Eqs.~(\ref{1dmep}) and
(\ref{1dmeq}), the master equations for $p_k(t)$ and $q_k^n(t)$ are
\begin{eqnarray} 
\label{mulmep}
p_k(t+1)&=&p_{k-1}\big[q_{k-1}^0
   +\sum_{n=1}^M q_{k-1}^n(1-{T}_n)\big]\\
&&\nonumber\\
\label{mulmeq}
q_k^n(t+1) &=& q_k^n(1-p_k{T}_n)
   +p_k q_k^{n+1}{T}_{n+1}.
\end{eqnarray}
In Eq.~(\ref{mulmep}), $q_{k-1}^{0}(t)$ accounts for the case that the pore
$k-1$ has zero capacity.  Other terms in Eq.~(\ref{mulmep}) correspond to
cases when the capacity is different from 0, with $(1-{T}_n)$ the survival
probability for each case. In Eq.~(\ref{mulmeq}) $q_k^{n}(1-p_k T_n)$ is the
probability that the pore with capacity $n$ does not trap a particle, and
$p_k q_k^{n+1}{T}_{n+1}$ is the probability that the capacity decreases from
$n+1$ to $n$ by a particle trapping event. Hence the last term is absent when
$n=M$.

We simplify Eqs.~(\ref{mulmep}) and (\ref{mulmeq}) by introducing the {\em
  average capacity} of a pore at position $k$,
\begin{equation}
\label{Qdef}
Q_k(t)\equiv \sum_{n=1}^M nq_k^{n}(t).
\end{equation}
This gives the average number of sites still available for trapping in the
pore. Now by multiplying Eq.~(\ref{mulmeq}) by $n$, summing from 1 to $M$,
and using $\sum_{n=1}^M q_k^{n}(t)=1-q_k^0(t)$, we obtain
\begin{eqnarray} 
\label{mulmep1}
p_k(t+1)&=&p_{k-1}\bigl(1-{\gamma\over M} Q_{k-1}\bigr)\\
\label{mulmeq1}
Q_k(t+1)&=&Q_k\bigl(1-{\gamma\over M} p_k\bigr).
\end{eqnarray} 
These are identical in form to Eqs.~(\ref{1dmep}) and (\ref{1dmeq}), so the
same steady state analysis applies.  We transform to a co-moving frame and
take the continuum approximation to reduce the rate equations to
Eqs.~(\ref{1dpwave}) and (\ref{1dqwave}) with $\gamma\to \gamma/M$ and 
$q\to Q$.  The boundary conditions are also the same as in the case of
single-capacity pores, except $Q(\xi\to\infty)=M$.  Combining these results
give
\begin{equation}
\label{mulvw}
v={v_0\over 1+M}\,,\qquad w(\infty)\sim{v_0 M\over \gamma(1+M)}.
\end{equation}
Notice that for $M=1$, Eq.~(\ref{mulvw}) reduces to the single capacity case,
while for $M\to\infty$, $v\to 0$.  This means that there is no steady state
for the case of infinite capacity pores.

We can generalize the symmetry argument given in the single capacity case to
find the propagation velocity in Eq.~(\ref{mulvw}). At the input, the flux of
invaders moving with the carrier fluid is equal to $1\times v_0$. Similarly,
in the reference frame moving with velocity $v_0$, the flux of defenders is
$M\times v_0$, while the invaders are at rest.  Because one invader
annihilates with one defender, the two particles are kinetically
indistinguishable.  Therefore, if a particle flux of $1\times v_0$ results in
a front moving with velocity $v$, the front velocity produced by a flux of
$M\times v_0$ should be $Mv$, which, in turn, equals $v_0-v$ in the moving
reference frame.  By this equivalence, Eq.~(\ref{mulvw}) immediately follows.

In the limit of perfect mixing, a particle encounters all traps in the pore.
The overall trapping probability with $n$ open traps is then
${T}_n=1-(1-\gamma)^n$. In the limit of small $\gamma$, ${T}_n\simeq\gamma
n$, thus the analysis is exactly the same as in the poor mixing case except
without the factor $1/M$ in ${T}_n$.  The propagation velocity of the front
is the same as in Eq.~(\ref{mulvw}), since this velocity is independent of
trapping mechanism, while the width varies as $w\sim v_0/\gamma(1+M)$.
Notice that $w$ is a decreasing function of $M$.  This arises because a
particle must survive all the traps in a pore before advancing to the next
pore.  Finally, for a mixing mechanism which is intermediate between the two
limits of perfect and poor mixing, the propagation velocity will be
$v_0/(1+M)$, while the width of the front will lie between the limiting
values of $v_0/\gamma(1+M)$ and $v_0M/\gamma(1+M)$.


\section{The bubble model}\label{bubble}

\begin{figure}
\narrowtext
\hskip.5cm\epsfxsize=62mm\epsfbox{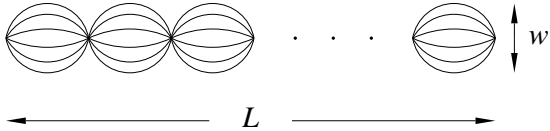} \vskip.4cm
\caption{The Bubble model consists of $L$ bubbles connected in series,
each bubble with $w$ tubes.
}
\label{fig4}
\end{figure}

We now study the bubble model as a logical next step towards understanding
infiltration in porous media.  The bubble model was introduced to account for
the breaking of fibers \cite{bubbles}, extremal voltages in resistor networks
\cite{kahng}, and later to filtration kinetics \cite{datta}.  The bubble
model consists of $L$ ``bubbles'' in series, each of which is a parallel bundle
of $w$ tubes, with each tube representing a pore (Fig.~\ref{fig4}).  A bubble
can be viewed as a single layer of parallel bonds in a lattice with all the
ends ``shorted''.  This model has multiple paths, as in real porous media,
and is sufficiently simple to be amenable to analytic study.  A useful
feature of the bubble model is that for straining dominated filtration, this
model predicts similar behavior to that of lattice networks \cite{datta}.

We choose the tube radii in the bubble model from the Hertz distribution
\begin{equation}
\label{hertz}
f(r)=2\alpha re^{-\alpha r^2},
\end{equation}
where $\alpha^{-1/2}$ is the characteristic pore radius.  This form is often
seen in experimental pore size measurements \cite{thomas} and has been used
for modeling the pore size distribution in filters \cite{sahimi,datta}.  For
simplicity we assume identical tube lengths and measure downstream distance
in units of the tube length, which is set equal to 1.  We also assume that
the flow rate in a tube of radius $r$ is proportional to $-r^\mu\nabla P$,
where $\nabla P$ is the pressure gradient along the tube and $\mu$ depends on
the nature of the flow, with $\mu=4$ corresponding to Poiseuille flow and
$\mu=2$ to Euler flow.  Perfect mixing is assumed at each node.  A particle
chooses a tube in the next downstream bubble according to {\em flow induced
  probability} $\Phi(r)$, 
\begin{equation}
\label{zeta}
\Phi(r)={r^\mu \over \int\!dr'\, f(r')\,r'^\mu},
\end{equation}
in which the probability of choosing an outgoing tube of radius $r$ is
proportional to the flow rate into the tube, $r^\mu$ \cite{datta,rege1}.

Since tubes of different flow velocities give the dominant mechanism for
dispersion, the radial dependence of the local flow velocity in a tube
(Taylor dispersion) is ignored.  Thus we assume that a particle moves with
the average flow velocity $v(r)$ along the tube.  We now investigate the
hydrodynamic dispersion of passive Brownian particles which are carried by
the background fluid in the bubble model, in the absence of any trapping.
This will provide the concepts and tools necessary to understand infiltration
in the bubble model.


\subsection{Hydrodynamic dispersion in the bubble model}\label{hydro}

In the large $w$ limit, each bubble is nearly identical, and we can regard
the particle motion as a directed random walk in which the average residence
time $\tau_k$ in bubble $k$ ($k=0, 1, 2,\ldots$) is a random variable whose
distribution $R(\tau)$ is related to the flow induced entrance probability
$\Phi(r)$ and the radius distribution $f(r)$.  This random walk description
of the continuous particle motion introduces an additional stochasticity into
the system.  However, we will show below that this only modifies the
hydrodynamic dispersion coefficient by an overall multiplicative factor.

The master equations for $p_k(t)$, the probability that there is a particle
in the $k^{\rm th}$ bubble at time $t$, are
\begin{equation}
\label{bubbleme}
{dp_0\over dt}=\rho\phi -{p_0\over \tau_0},\quad
{dp_k\over dt}= {p_{k-1}\over\tau_{k-1}}-{p_k\over\tau_k}.
\end{equation}
Here $\rho$ is the initial particle number concentration, $\phi$ is the 
(constant) flow rate, and the initial condition is $p_k(0)=0$ for all $k$.  
Since the flow rate does not change in infiltration, constant pressure drop 
and constant flow rate conditions are equivalent.

The particle transport properties can be obtained in terms of the residence
time distribution $R(\tau)$, namely, the probability that a particle spends a
time $\tau$ in a bubble.  This residence time distribution is related to
microscopic distributions by
\begin{equation}
\label{ftau}
 R(\tau)=\int_0^\infty\!dr \,\Phi(r)\,f(r) \, \delta(\tau-{1\over v(r)}),
\end{equation}
where $\delta(x)$ is the Dirac delta function.  Since the flow rate into a
tube of radius $r$ is $\phi \Phi(r)$, the average flow velocity $v(r)=\phi
\Phi(r)/\pi r^2$.  Using this together with Eq.~(\ref{hertz}) for $f(r)$, we
obtain the first two moments of $\tau$
\begin{eqnarray}
\label{tau1}
\langle{\tau}\rangle&=&\int_0^\infty \tau' R(\tau')d\tau'
           ={\pi\over \phi\alpha} \equiv {V\over \phi}\\
\label{tau2}
\langle{\tau^2}\rangle&=&
   \langle{\tau}\rangle^2\Gamma(1+{\mu\over2})\Gamma(3-{\mu\over2}),
\end{eqnarray}
where $V=\langle\pi r^2\rangle$ is the average tube volume (recall that the
tube length is fixed to be 1) and $\Gamma(x)$ is the gamma function.

We solve Eq.~(\ref{bubbleme}) in the Appendix by the Laplace transform
technique.  From this solution, the average propagation velocity and the
width of the front are
\begin{eqnarray}
\label{hydrov}
 v_0&\simeq& {1\over\langle\tau\rangle}\\
\label{hydrow}
 w&\simeq& \left\{{t\over\langle{\tau}\rangle}\left[
  2\Gamma(1+{\mu\over2})\Gamma(3-{\mu\over2})-1\right]\right\}^{1\over2}\equiv
  (D_\parallel t)^{1\over2}.
\end{eqnarray}

Thus we see that the dispersion coefficient is proportional to the average
flow velocity $\langle\tau\rangle^{-1}$.  When $\mu=2$ (Euler flow), the flow
velocities in all the tubes are identical and there should be no dispersion.
However, Eq.~(\ref{hydrow}) gives a nonzero dispersion coefficient.  As
mentioned above, this arises from the stochasticity of the random walk
picture for the particle motion.  For the practically relevant case of
$\mu\simeq4$, the effect of this stochasticity is only to change the
dispersion coefficient by a factor of order unity.


\subsection{Infiltration in the bubble model}\label{bubbleinf}

To describe infiltration in the bubble model, we need to specify the particle
motion, the tube capacities, and particle trapping in a tube.  For the
particle motion we again assume that a particle chooses a tube according to
flow induced probability and then advances with the average flow velocity
$v(r)$ of this tube.  The capacity of a tube is proportional to its inner
surface area, which is proportional to the tube radius, since all tubes have
the same length.  Last, the overall trapping probability of a tube is equal
to the microscopic trapping probability $\gamma$ multiplied by the fraction
of open traps in a tube (Sec.~\ref{1dmulti}).

To simulate this process efficiently we propagate the probability
distribution function (PDF) of the suspended particles rather than simulating
the motion of individual particles\cite{redner,pdf}.  The PDF propagation
therefore provides the exact distribution of particle positions and tube
capacities for a single realization of tube radii.  Conceptually, the PDF
algorithm is equivalent to an exact integration of the master equations.

\begin{figure}
\narrowtext
\hskip1cm\epsfxsize=55mm\epsfbox{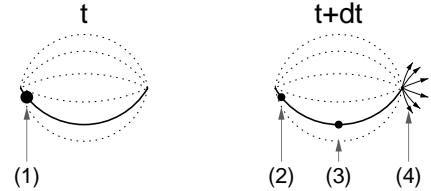} \vskip.4cm
\caption{Propagation of the PDF in tube $\beta$ at bubble $k$.
  (1) Initial probability element $p_k^\beta(t)$. (2) Fraction remaining.
  (3) Fraction trapped.  (4) Fraction advancing.  This last element enters
   the next bubble and is then immediately split among the
   tubes according to the flow induced probabilities.}
\label{fig5}
\end{figure}

To implement the PDF propagation, we define
\begin{itemize}
\item $p_k^\beta(t)$: The probability that there is a particle at the {\em
    entrance}\/ of tube $\beta$ in bubble $k$ ($\beta=1,2,\ldots, w$,
  $k=0,1,2,\ldots$).
\item $c_k^\beta(t)$: The capacity of tube $\beta$ in bubble $k$.
\end{itemize}
Since particles generally have different velocities, their positions could be
anywhere within a tube.  We simplify this by forcing particles to {\em
  always} be at the tube entrance by adjusting the time unit and the PDF
propagation so that the {\em average} particle position is at the correct
location along the tube, as illustrated in Fig.~\ref{fig5}.

To construct the particle motion, let us temporarily disregard particle
trapping.  We set the time increment to be $\delta t=1/v_{\rm max}$, where
$v_{\rm max}$ is the maximum flow velocity among all tubes.  In a time
$\delta t$, a particle at the entrance of the fastest tube should traverse
the entire tube length which is equal to 1.  We then let a particle in a
slower tube, with velocity $v_k^\beta<v_{\rm max}$, travel a distance $1$
with probability $u_k^\beta\equiv{v_k^\beta/v_{\rm max}}$, or remain fixed 
with probability $1-u_k^\beta$. One can regard
$u_k^\beta$ as a normalized flow velocity.  By construction, such a
particle travels the correct average distance in time $\delta t$, $1\cdot
u_k^\beta +0\cdot(1-u_k^\beta )= {v_k^\beta/ v_{\rm max}}= v_k^\beta
\delta t$.

Let us now recast this random walk into a probability propagation algorithm.
Consider an element of the PDF which is at the junction before the $k^{\rm
  th}$ bubble.  Before any particle motion occurs, we split this probability
element among the downstream bonds in this bubble according to the flow
induced probability at the tube entrance.  We can view the probability
element as advancing infinitesimally into each bond, as indicated on the left
side of Fig.~\ref{fig5}.  Once this initial tube assignment is made, the
probability element remains within its assigned tube until it reaches the
next junction.

Now consider the motion of a probability element which has just entered a
particular bond.  After a time $\delta t$, a fraction $u_k^\beta$ of the PDF
is advanced to the next bubble, while a fraction $1-u_k^\beta$ remains fixed
at the entrance to bond $\beta$.

Due to the filtration, a fraction of the flowing PDF becomes trapped in tube
$\beta$ in the $k^{\rm th}$ bubble at a rate which is proportional to the
tube capacity $c_k^\beta(t)$.  The overall trapping probability of this tube
is therefore ${T}_k^\beta=\gamma c_k^\beta(t)/ c_k^\beta(0)$.  
After trapping has occurred, the tube capacity is decremented according to
the following prescription.  When one unit of PDF (equivalent to one
particle) gets trapped, we define the bond capacity to be decreased by
$\Delta$.  Therefore $\Delta$ is just the surface area covered by one
particle.  Correspondingly, $c_k^\beta(0)/\Delta$ equals
the number of particles the tube can accommodate.

Our algorithm for propagating an element of probability at the entrance to
bond $\beta$ in the $k^{\rm th}$ bubble over a time $\delta t$ therefore 
consists of the following steps (Fig.~\ref{fig5}):
\begin{itemize}
\item Fraction of PDF remaining at the start: $p_k^\beta(1-u_k^\beta )$.
\item Fraction trapped in tube $\beta$: 
      $p_k^\beta u_k^\beta{T}_k^\beta$.
\item Fraction advancing to the next junction:

  $p_k^\beta u_k^\beta (1-{T}_k^\beta)$
\item Capacity change of the tube by trapping: 

     $-\Delta\times [p_k^\beta
        u_k^\beta{T}_k^\beta] $.
\end{itemize}
The rate equations which account for these steps are:
\begin{eqnarray}
\label{bubblepdf}
p_k^\beta(t+\delta t)&=& 
   p_k^\beta(1- u_k^\beta )\nonumber \\
   &&+ \Phi_k^\beta\sum_{\beta'=1}^{w}p_{k-1}^{\beta'}
   u_{k-1}^{\beta'} (1-{T}_{k-1}^{\beta'})\\
&&\nonumber\\
\label{bubblecap}
c_k^\beta(t+\delta t)&=&c_k^\beta
    -\Delta p_k^\beta u_k^\beta{T}_k^\beta,
\end{eqnarray}
The first term on the right hand side of Eq.~(\ref{bubblepdf}) is the
fraction of probability that does not move, and the second term is
contribution from elements of probability which has moved from the previous
site.  The flow induced probability $\Phi_k^\beta$ in Eq.~(\ref{bubblepdf})
accounts for the fraction of PDF which enters into tube $\beta$.

To test this approach, we set $\gamma=0$ (no trapping) in the above rate
equations and simulate the PDF propagation.  By this method, we find
a traveling front whose basic properties coincide with the
hydrodynamic dispersion results given by Eqs.~(\ref{hydrov}) and
(\ref{hydrow}).

It is also worth mentioning that our PDF algorithm can be generalized to
allow for hopping a distance which is a fraction of the tube length.  In
this manner one can account for different longitudinal flow velocities at
different radial positions within a tube (Taylor dispersion \cite{brenner}).
In the limit of an infinitesimal hopping distance, continuous particle motion
is reproduced by the PDF algorithm.  Unfortunately, the gain in having a more
accurate description of the motion is offset by the complexity of the
algorithm and large increase in the computation time.


\subsection{Asymptotic behavior}\label{bubbleanal}

To obtain the average invasion front profile over the tubes in each bubble,
we define the bubble-average quantities
$P_k(t)\equiv {1\over w}\sum_\beta p_k^\beta(t)$ and
$Q_k(t)\equiv {1\over w}\sum_\beta c_k^\beta(t)$. We first derive the
invasion front velocity via the symmetry argument of Sec.~\ref{1dmulti}.
A rigorous derivation of the front velocity from the master equations for 
$P_k(t)$ and $Q_k(t)$ is given in \cite{thesis}.

The carrier fluid is moving with velocity $v_0$ (Eq.~(\ref{hydrov})) and the
input flux of invaders per tube is equal to $v_0\rho V$, since $\rho V$ is
equal to the number of invader particles per tube volume in the input fluid.
On the other hand, in a reference frame moving with velocity $v_0$, the input
flux of defenders is equal to $v_0M$ where $M\equiv\langle r\rangle/\Delta$
is the average initial number of invaders a tube can accommodate. Following
the argument in Sec.~\ref{1dmulti},
we find the front velocities $v$ and $v_0-v$ in the two reference frames
are related by $Mv=\rho V(v_0-v)$, yielding
\begin{equation}
\label{bubblev}
{v\over v_0}=\left(1+ {M \over\rho V}\right)^{-1}.
\end{equation}

Good filter performance means that the
breakthrough time is long or, equivalently, that the propagation velocity is
slow.  Eq.~(\ref{bubblev}) implies that the propagation velocity can be made
small by increasing the capacity of a pore, or by decreasing
either the filter grain size or the input particle concentration.  Notice
that neither the reaction strength $\gamma$ nor the nature of the flow
(through $v(r)\sim r^{\mu-2}$) affect this propagation velocity.

We now study the asymptotic density profiles. Instead of working directly
with the averaged quantities $P_k(t)$ and $Q_k(t)$, we first focus on the
behavior of a single tube of radius $r$, since tubes with the same radius in
a bubble have identical time dependence. The asymptotic profiles can be
obtained after averaging over the distribution of tube radii.  Therefore, we
label tubes according to their radii instead of the index $\beta$.  We denote
$p_k(t;r)$ and $c_k(t;r)$ as the PDF and capacity of a tube of radius $r$ in
the $k^{\rm th}$ bubble. 

Let us first focus on the PDF profile in the invaded region.  Here, traps
are mostly unoccupied, so that the tube capacity $c_k(t;r)$ is approximately
equal to its initial value $c_k(0;r)$, which is proportional to the 
tube radius and we set it equal to $r$. The arbitrariness in the unit of 
capacity can be controlled by the magnitude of the parameter $\Delta$. 
Then Eq.~(\ref{bubblepdf}) becomes, 
\begin{eqnarray}
\label{attack}
 p_k(t+\delta t;r)&\simeq& p_k(t;r)\left(1-u(r)\right)\nonumber\\
+&&\hskip-.4cm(1-\gamma)\Phi(r)\int\! dr' f(r')p_{k-1}(t;r')u(r'),
\end{eqnarray}
where the integration over $r'$ replaces the summation over the tube index,
The flow induced probability $\Phi(r)$ and normalized velocity $u(r)$ are
independent of the downstream position $k$ because these only depend on the 
radius of a tube.

The equation for $P(x,t)$ can be obtained by multiplying Eq.~(\ref{attack})
by $f(r)$ and integrating over $r$.  As in Sec.~\ref{1d}, we take $\gamma\ll
1$ and consider the continuum limit.  If we redefine the length of the bubble
from 1 to $\delta x$, $\delta t$ becomes $\delta x/v_{\rm max}$. Integrating
Eq.~(\ref{attack}) and expanding in $\delta t$ and $\delta x$ yields
\begin{equation} 
\label{attack0}
\delta t\partial_t P(x,t)=-\int\!dr f(r)u(r)
       [\delta x\partial_x p(x,t;r)+\gamma p(x,t;r)]
\end{equation}
Here, we use $\int\!drf(r)\Phi(r)=1$. Dividing
Eq.~(\ref{attack0}) by $\delta t$ changes $u(r)$ back to $v(r)$.  
After redefining
$\gamma\rightarrow\gamma/\delta x$ as before, we obtain
\begin{equation}
\label{attack1}
\partial_t P\simeq-\int\!dr f(r)v(r)\left[\partial_xp(x,t;r)
+\gamma p(x,t;r)\right].
\end{equation}
In the steady state co-moving frame, Eq.~(\ref{attack1}) becomes
\begin{equation}
\label{attack2}
v\partial_\xi P\simeq\int\!drf(r)v(r)
     \left[\partial_\xi p(\xi;r)+\gamma p(\xi;r)\right].
\end{equation}

Since only a small number of particles have entered the invaded region, the
density of moving particles is approximately proportional to $r^\mu$, and we
introduce the ansatz $p(\xi;r)=r^\mu g(\xi)$ to factorize the PDF.  
In order to calculate the dominant contribution from the integral 
in Eq.~(\ref{attack2}), we 
substitute $v(r)=v_0+\delta v(r)$, where $\delta v(r)$ is the deviation from
the average carrier fluid velocity $v_0$.  Since $\delta v(r)$ has zero mean,
the dominant contribution to the integral over $v(r)$ in Eq.~(\ref{attack2})
comes from the constant part $v_0$.  Using these approximations in
Eq.~(\ref{attack2}), and using $P(\xi)=\langle r^\mu\rangle g(\xi)$, we find
\begin{equation}
\label{attack3}
v g'\simeq v_0(g'+\gamma g).
\end{equation}
Since $v<v_0$, we find $g(\xi)\sim \exp\left[-v_0\gamma\xi/(v_0-v)\right]$.
Hence the profile of free particles in the invaded region $P(\xi)$ decays
exponentially in $\xi$, with a characteristic decay length which has the same
$1/\gamma$ dependence as in the 1D model.

Let us now turn to the analysis of the capacity profile in the tail region. 
In terms of $p(x,t;r)$ and $c(x,t;r)$, Eq.~(\ref{bubblecap}) becomes
\begin{equation}
\label{tail}
\partial_t c(x,t;r)=-\Delta\gamma p(x,t;r)v(r){c(x,t;r)\over r}.
\end{equation}
Since there is negligible trapping in the tail region, the particle motion
follows that of the carrier fluid.  Thus $p(\xi;r)\simeq \pi r^2\rho$, where
$\pi r^2$ is the tube volume, and the flow velocity is $v(r)=\phi\Phi(r)/\pi
r^2=\phi r^{\mu-2}/\pi \langle r^\mu\rangle$.  Substituting these in
Eq.~(\ref{tail}) and transforming into the co-moving frame gives
\begin{equation}
\label{tail1}
\partial_\xi c(\xi;r) \simeq s r^{\mu-1}c(\xi;r),
\end{equation}
where $s\equiv\Delta\gamma\rho\phi/v\langle r^\mu\rangle$ denotes the 
strength of the particle trapping reaction.  We now
integrate Eq.~(\ref{tail1}) from $-\xi_1\ll 0$ to $\xi_2\approx0$ and use the
boundary condition $c(\xi_2;r)\simeq r$ to obtain
\begin{equation}
\label{tail2}
c(\xi;r)\simeq r e^{-s r^{\mu-1}|\xi|},
\end{equation}
where we drop the subscript of $\xi_1$.

Finally, the average bond capacity as a function of position with respect to
the front, $Q(\xi)$, is
\begin{equation}
\label{tail3}
Q(\xi)=\int\!dr\,f(r)c(\xi;r) \simeq 2\alpha \int\! dr\, 
r^2 e^{-s r^{\mu-1}|\xi|-\alpha r^2}.
\end{equation}
For large $|\xi|$, the integral is dominated by the smallest tubes and the
initial distribution of tube radii is irrelevant in the tail region.  Hence
the factor $\alpha r^2$ in the exponential can be ignored.  Performing the
resulting integration gives
\begin{equation}
\label{tail4}
Q(\xi)\sim{(s|\xi|)^{-{3\over\mu-1}}\over(\mu-1)}
\Gamma({3\over\mu-1})\propto (\gamma|\xi|)^{-\lambda},
\end{equation}
where the last relation serves to define the {\em profile exponent}
$\lambda$.  This is one of our primary results. Correspondingly, the PDF 
in the tail region will approach its initial value with the same power law.

The existence of the power law tail in the capacity profile stems from the 
fact that the flow rate is not affected by trapping.  Thus when large pores 
are ``used up'', the fluid still predominantly flows through these pores, 
leading to a substantial unused capacity in the smaller tubes.  It is these 
unused smaller tubes which contribute substantially to the capacity profile 
in the tail region. 
This mechanism is quite general and only depends weakly on the form of the
radius distribution.  For example, for a uniform distribution in the range
$r=(0,1)$, we obtain $\lambda=2/(\mu-1)$.  However, if there is a finite 
lower cutoff in the radius distribution, the PDF will have an asymptotic 
exponential tail.

It is interesting to note that the density profile has different dependence
on $\gamma$ in the invaded and tail regions.  From Eq.~(\ref{tail4}),
the density profile contains an
overall factor $\gamma^{-\lambda}$.  Thus $\gamma$ typically does not
appear as an overall scale factor of the entire profile, as in the invaded
region.  However, for the practically relevant case of $\mu=4$, the exponent
in Eq.~(\ref{tail4}) is equal to 1, and $\gamma$ becomes the overall scale
factor of the profile.


\subsection{Numerical results}\label{bubbledata}

In our numerical simulations, we set the input particle flux per tube
$\rho\phi$ equal to 1, which means that $w$ units of PDF are injected into
the system at every time step. This can be achieved by choosing
$\rho=1/\pi$ and $\phi=\pi$, which also makes $v_0=\alpha$
(Eqs.~(\ref{tau1}) and (\ref{hydrov})).

\begin{figure}
\narrowtext\epsfxsize=75mm\epsfysize=54mm\hskip.5cm
\epsfbox{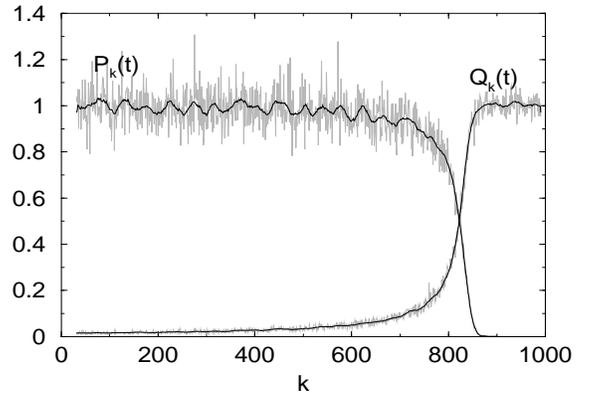} 
\vskip.4cm
\caption{Normalized profiles for a single realization. 
  Parameters used are $\mu=4$, $\alpha=1$, $\gamma=0.1$, and $\Delta=0.4$.
  Gray lines are raw data, black lines are smoothed.  }
\label{fig6}
\end{figure}

We applied the PDF propagation of Eqs.~(\ref{bubblepdf}) and
(\ref{bubblecap}) to a system of size $w\times L= 200\times1024$.  Due to the
exact nature of the PDF algorithm, a single realization provides good quality
data for $w=200$ tubes.  A system length of $L=1024$ is sufficiently long to
give the continuum functional form of the profiles.  All the data shown 
below are results of single realization of tubes including the network 
simulation in the next section. The simulation is stopped before the front 
exits the system.

{\it Density Profiles}. Fig.~\ref{fig6} shows typical particle and tube
capacity profiles.  There are strong bubble-to-bubble fluctuations and some
type of smoothing procedure is necessary.  We use the Savitzky-Golay
smoothing technique, which approximates
successive windows of data points 
to a $4^{\rm th}$ order polynomial (solid lines in the figure)
\cite{numrecipe}.  This technique is superior to local averaging because
Savitzky-Golay smoothing can faithfully follow rapid changes in the profile,
as can be seen in Fig.~\ref{fig6}.  This smoothing is also useful in
estimating the exponents.  Since logarithm of the profile in the tail region
amplifies the fluctuation in nonlinear way, slopes of the raw data in
Fig.~\ref{fig_tail1} are larger than those from the smoothed data, and
differs from the predicted value of the profile exponent $\lambda$.

\begin{figure}
\narrowtext
\hskip.1cm\epsfxsize=80mm
\epsfbox{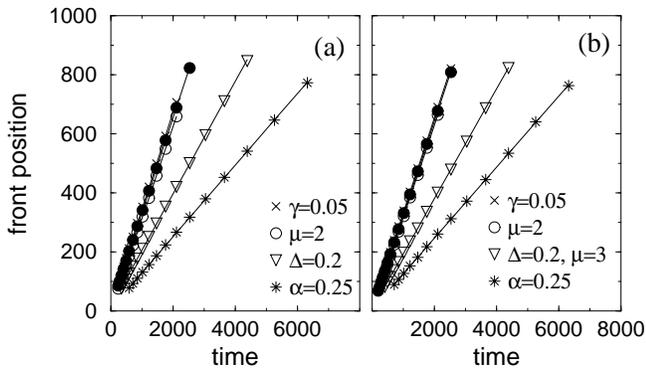}
\vskip.4cm
\caption{Front position vs.\ time for different microscopic parameters for
  (a) the bubble model and (b) the square lattice.  
  Filled circles are for the same parameters as those in Fig.~\ref{fig6}.
  Other data modify these parameters as indicated in the
  legend.  The straight lines are linear fits to the data with slopes as 
  (a) 0.33, 0.32, 0.31, 0.19, 0.12, and (b) 0.32, 0.32, 0.31, 0.19, 0.12,
  from top to bottom.  The corresponding velocities predicted by 
  Eq.~(\ref{vel}) are 0.311, 0.311, 0.311, 0.184, and 0.119. }
\label{fig_vel}
\end{figure}

{\it Front Velocity}.  Fig.~\ref{fig_vel}(a) shows the front position, defined as
the point where $P_k(t)$ is half of its saturation value, versus time.
Notice that a constant front propagation velocity sets in almost immediately.
With $\rho=1/\pi$ and $\phi=\pi$, Eq.~(\ref{bubblev}) gives
\begin{equation}
\label{vel}
{v\over v_0}={1\over 1+{\sqrt{\pi\alpha}/ 2\Delta}}.
\end{equation}
The slopes in Fig.~\ref{fig_vel}(a) agree well with Eq.~(\ref{vel}).  Notice
that the propagation velocity does not depend on the reaction strength
$\gamma$ nor the exponent $\mu$ in the radius dependence of the velocity.

\vskip.5cm {\it Tail Profile}.  Fig.~\ref{fig_tail1}(a) shows the tube
capacity profile $Q(\xi)$ in the tail region as a function of the distance
$|\xi|$ ($\xi<0$) from the front on a double logarithmic scale.  The plot
becomes straight for large $|\xi|$ and the slope in this region corresponds
to the exponent $\lambda=3/(\mu-1)$ predicted by Eq.~(\ref{tail4}).  For the
uniform distribution on $(0,1)$, we predicted the profile exponent to be
$\lambda=2/(\mu-1)$.  For $\mu=4$, the exponent value of 2/3 agrees well with
our simulations (Fig.~\ref{fig9}).  However, for a radius distribution with 
a lower size cutoff, we expect an exponential density profile 
(inset to Fig.~\ref{fig9}).

\begin{figure}
\narrowtext
\hskip.1cm\epsfxsize=82mm\epsfbox{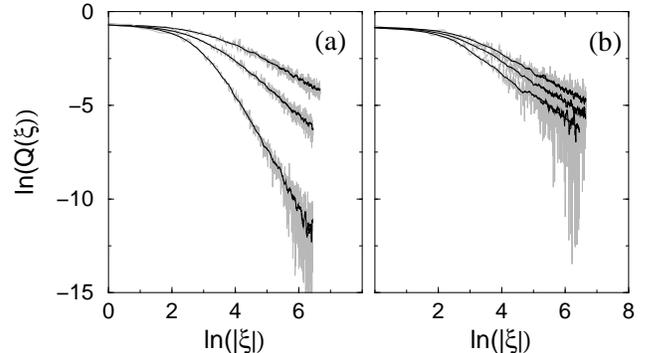}\vskip.4cm
\caption{Tube capacity profile in the tail region on a double
  logarithmic scale for $\mu=4$, 3, and 2 (top to bottom).  Other parameters
  are the same as in Fig.~\ref{fig6}.  (a) Bubble model: the slopes of the
  data in straight region are 0.97 (1), 1.49 (1.5), and 3.17 (3).  The
  numbers in parenthesis are the prediction $3/(\mu-1)$.  (b) Square lattice:
  Solid lines are smoothed data.  The thick straight lines are linear fits,
  with slopes 0.95, 1.07, and 1.28.}
\label{fig_tail1}
\end{figure}

\begin{figure}
\narrowtext\epsfxsize=75mm
\hskip.5cm
\epsfbox{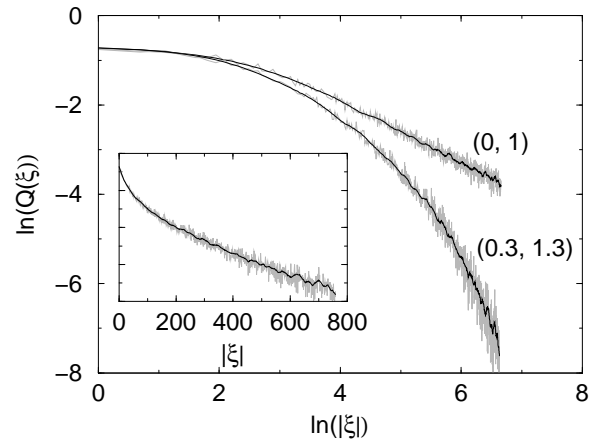} \vskip.4cm
\caption{Tube capacity profile on a double logarithmic scale for 
  a uniform radius distribution on $(a,b)$.  Parameters other than the radius
  distribution are the same as in Fig.~\ref{fig6}.  The slope of the straight
  region for $(a,b)=(0,1)$ is 0.63 (predicted value 2/3).  Inset: log-normal
  plot when $(a,b)=(0.3,1.3)$.}
\label{fig9}
\end{figure}

\begin{figure}
\narrowtext
\hskip-.2cm\epsfxsize=82mm\epsfbox{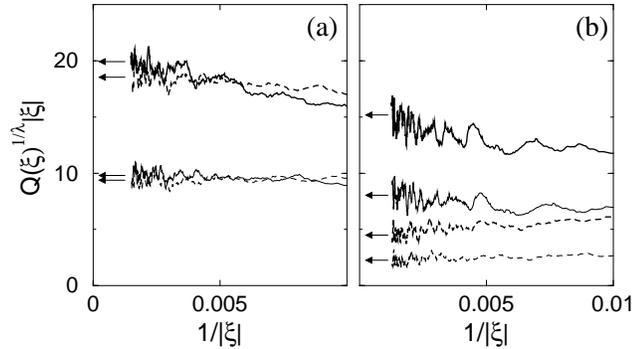}\vskip.4cm
\caption{Dependence of the capacity profile in the tail region on $\gamma$.
  Only smoothed data are shown. Solid curves: $\mu=4$, dashed curves:
  $\mu=3$. Thick curves: $\gamma=0.05$, thin curves: $\gamma=0.1$.
  Other parameters are the same as in Fig.~\ref{fig6}.
  (a) Bubble model: 
  The value $\lambda=3/(\mu-1)$ is used. 
  (b) Square lattice: From Fig.~\ref{fig_tail1}(b), $\lambda=0.95$ 
  for $\mu=4$ and 1.07 for $\mu=3$ are used. 
  Notice that the values of abscissa are proportional to $1/\gamma$
  from Eq.~(\ref{tail4}). Arrows are guides to the eye. }
\label{fig_tail2}
\end{figure}

As we also discussed in Sec.~\ref{bubbleanal}, the amplitude of the density
profile in the tail region typically has a power-law dependence on $\gamma$.
For a Hertz distribution of particle radii, this amplitude should be
proportional to $\gamma^{-\lambda}$ according to Eq.~(\ref{tail4}).  Thus
Fig.~\ref{fig_tail2}(a) shows $Q(\xi)^{1/\lambda}|\xi|$ versus $1/|\xi|$ for
$\mu=4$ and $\mu=3$.  Values of the abscissa should be proportional to
$1/\gamma$, which is indeed the case.

\begin{figure}
\narrowtext
\hskip.2cm
\epsfxsize=80mm\epsfbox{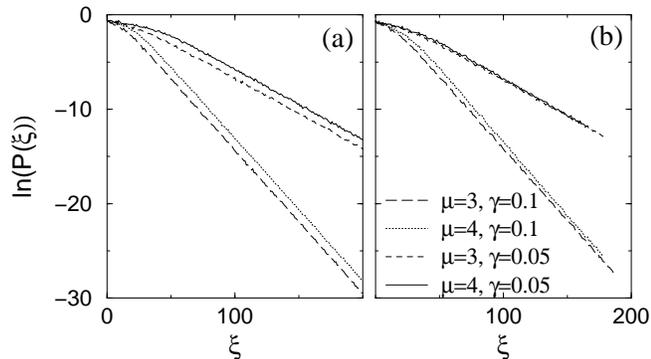}
\vskip.4cm
\caption{Log-normal plot of the invaded region. The same data sets as
  in Fig.~\ref{fig_tail2} are used. The slopes of the straight regions are,
  from left to right, (a) bubble model: 0.152, 0.150, 0.0744, 0.0748, and 
  (b) square lattice: 0.152, 0.155, 0.0756, 0.0773. The corresponding 
  values of the slope from $v_0\gamma/(v_0-v)$ are 0.145 and 0.0726.} 
\label{fig_invade}
\end{figure}

{\it Invader Profile}. Fig.~\ref{fig_invade}(a) is a log-normal plot of the
invader profile $P(\xi)$ (raw data) versus $\xi$.  The slopes of the two data
sets are in excellent agreement with the predicted value $v_0\gamma/(v_0-v)$
from Eq.~(\ref{attack3}).  In the invaded region, since there are few
particles present, the corresponding PDF monotonically decreases and its
fluctuation is significantly smaller compared to the tail region.

\section{Square lattice network}\label{network}

We now consider infiltration on a square lattice network of tubes.  Here,
local mixing at tube junctions occurs as opposed to the mean-field-like
mixing in the bubble model.  Nevertheless, many of our predictions from the
bubble model continue to be valid for the lattice network.  For example,
we expect that the propagation velocity given by Eq.~(\ref{bubblev}) will
continue to hold in the lattice network because it is determined {\em only}
by the boundary conditions of the particle flux and initial filter capacity.
We also find numerically that the network model has both the same exponential
invader profile and the power law capacity profile as in the bubble model,
although the values of the amplitudes and decay exponents are different.
Overall, it appears that the bubble model provides an excellent account of
the numerical results from the lattice network.


\subsection{The model}\label{netsetup}

We study a square lattice of size $w\times L= 200\times 1024$ which is tilted
at $45^{\rm o}$.  A periodic boundary condition is imposed in the transverse
direction.  The tube radii are drawn from the Hertz distribution of
Eq.~(\ref{hertz}).  Notice that the bubble model would arise from this system
by merging together all sites at the same longitudinal position.  The overall
flow rate $\phi$ is set to $\pi$ and the particle density to $\rho=1/\pi$,
just as in the bubble model.

For a given set of tube radii, the flow field is calculated by using the
conjugate gradient method\cite{numrecipe} to solve the set of linear
algebraic equations for fluid conservation at each node.  The tolerance of
the computation is set so that the measured average PDF and the tube capacity
in the $k^{\rm th}$ layer are accurate to within 0.01\%.  After the flow
field is solved, we use the same PDF algorithm as in the bubble model to
track the motion of the suspended particles.  The new features due to the
lattice nature of the network are that tubes are only locally connected and
that the local flow direction is not always downstream.


\subsection{Numerical results}\label{netdata}

To facilitate comparisons, all our numerical results for the bubble model and
the square lattice are presented side-by-side.  Fig.~\ref{fig_vel} shows the
front position versus time for both the bubble model and the square lattice.
The square lattice results are in excellent agreement with the bubble model
prediction for the front velocity, Eq.~(\ref{vel}).  Similarly, the tube
capacity profile exhibits a power law tail
(Fig.~\ref{fig_tail1}(b)). However, the dependence of the decay exponent
$\lambda$ on $\mu$ is much weaker compared to the bubble model. As
$\mu$ decreases from 4 to 2, $\lambda$ increases
slowly from 0.95 to 1.28. 

To isolate the dependence of the capacity profile on the trapping probability
$\gamma$, Fig.~\ref{fig_tail2}(b) shows $Q(\xi)^{1/\lambda}|\xi|$ versus
$1/|\xi|$ in the tail region.  Here, the values of $\lambda$ used 
are obtained from Fig.~\ref{fig_tail1}(b).  Unlike the bubble model,
the overall amplitude has a relatively stronger dependence on $\mu$.
However, the dependence on $1/\gamma$ still holds even in the network case.
It seems that in the square lattice network, $\mu$ has more
effect on the amplitude of the tail than on the decay
exponent.

Lastly, Fig.~\ref{fig_invade}(b) shows the density profile of invaders in 
the invaded region. The slopes of these particles are almost identical to 
those of the bubble model. Currently we do not have clear explanation of 
this fact.
It seems that the network geometry does not 
affect the profile. In fact, the characteristic decay length,
$(v_0-v)/v_0\gamma$, of the profiles in Fig.~\ref{fig_invade} is 
the same as that of the corresponding 1D model in Sec.~\ref{1dmulti} 
where $\rho V$ invaders are injected with velocity $v_0$ into the chain of 
defenders of capacity $M$.


\section{Summary and Discussion}\label{discussion}
In this paper, we studied infiltration, in which suspended particles are
removed from a carrier fluid as the suspension passes through a porous
medium.  The trapping mechanism has a built-in saturation so that once all
available trapping sites are used up, subsequent particles can pass through
the medium freely.  The particles are assumed to be sufficiently small that
their trapping does not change the flow rate.  The basic dynamical properties
of this infiltration process are the density profile of the invader particles
and the capacity profile of the remaining active pores.  When the invader
profile reaches the end of the system, the output concentration of particles
quickly increases to a saturation level and the filter should be discarded.
Thus the features of this profile are important to understand the operating
characteristics of infiltration.

We have developed a series of discrete network models to describe the basic
characteristics of infiltration, starting with a one-dimensional model and
building up to the bubble model, which is a series array of parallel,
multiple-capacity tubes.  The advantage of these quasi-one-dimensional models
is that they remain relatively simple, even after incorporating local spatial
heterogeneity.  The bubble model, in particular, appears to capture many of
the quantitative features that we observed in numerical simulations of
infiltration on a square lattice tube network.  Our modeling is also
flexible, so that variations can be easily implemented for case-specific
situations.

Our main qualitative result is that basic dynamical features of the system,
including the value of the front propagation velocity, the exponential
profile of flowing particles in the invaded region, and the power law
capacity profile of pores in the tail region, are relatively insensitive to
microscopic details of the model.  We have also identified the basic
parameters which do affect quantitative features of the profiles.  It is
useful to summarize these results and to compare with experimental data, as
well as with predictions from previous studies based on the
reaction-diffusion equation approach \cite{luhrmann,putnam,tien}.

\begin{figure}
\narrowtext
\hskip.2cm
\epsfxsize=65mm\epsfbox{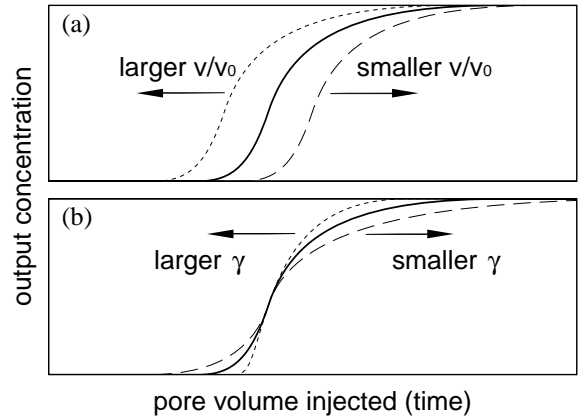}
\vskip.4cm
\caption{Qualitative dependence of the output concentration curve (a) on the 
  propagation velocity, and (b) on the reaction strength.}
\label{fig_discuss}
\end{figure}

{\em Invader Concentration at the Output}.  In typical
experiments, the invader concentration at the output is measured as a
function of time.  A slower propagation velocity shifts this output
concentration curve to a later time, as indicated in
Fig.~\ref{fig_discuss}(a).  Typically, the time unit is normalized by the
time for passive particles to pass through the system \cite{luhrmann,putnam}.
Hence, the amount of the time shift is determined by the ratio $v/v_0$ rather
than by $v$ itself.

A nice set of infiltration experiments, as well as an accompanying numerical
study of the reaction-diffusion equation were performed in \cite{putnam}.  In
these experiments, contaminant solutions with different values of the invader
particle diameter $d$, but with fixed mass concentration were used.  This
makes the corresponding number density $\rho$ of invaders in each solution
proportional to $d^{-3}$. Also, since the cross-sectional area of each
particle is proportional to $d^2$, the average initial tube capacity $M$
varies as $d^{-2}$. From Eq.~(\ref{bubblev}), we then have $v/v_0\sim
(1+d)^{-1}$.  Thus the output concentration curve shifts to a later time for
a solution with a larger value of $d$, which is consistent with the
experiment and numerical predictions in \cite{putnam}.

In another set of experiments, different electrolyte concentrations of the
carrier fluid were used.  This mainly affects the trapping rate $\gamma$.
For a larger trapping rate, the width of the output concentration curve,
namely, the time range over which the output concentration changes from zero
to its saturation value becomes narrower.  However, there is no shift in the
breakthrough time because the propagation velocity is independent of
$\gamma$.  These two features are illustrated in Fig.~\ref{fig_discuss}(b).
This behavior again qualitatively agrees with the experiment in 
\cite{putnam}.  

It would also be interesting to study the effect of different filter grain
sizes.  In our model this would be accomplished by changing the
characteristic parameter $\alpha$ of the pore size distribution.  In turn,
this affects the invader propagation velocity, and the output concentration
curve will shift in time accordingly.  However, since the microscopic
parameters we use in our modeling may be coupled with each other in
experimental situations, different sizes of filter grains or invader
particles may also affect, {\it e.\ g.}, the reaction strength $\gamma$.  We
can incorporate such a coupling effect by extending our model to deal with
these effects explicitly.  For example, we can adapt the microscopic models
of particle trapping on a single sphere or on a plane
\cite{probstein,rajagopalan}, to the tube geometry.  From such an approach,
we can express the reaction strength $\gamma$ as a function of the invader or
defender diameter.

{\em Tail of the Output Concentration Curve}. A slowly decaying tail in the
deviation of the output concentration from its asymptotic value is generally
observed in experiments \cite{luhrmann,putnam,tien}.  This observation is in
contrast to the empirical approaches, such as that given in \cite{putnam},
which gives an exponential profile for the whole time range.  Their
prediction agrees with experimental observations at early times but then
deviates at later times, implying that the output profile at later times is
not exponential.

A closely related approach, based on the study of a reaction-diffusion
equation, is presented in \cite{luhrmann}, along with experiments which
measure the output concentration.  Unlike \cite{putnam}, here the adsorption
rate depends on the local concentration of contaminants and thus is spatially
inhomogeneous.  A crossover from a rapid increase to a slowly decaying tail
of the output concentration was numerically predicted.  However, the
functional forms of these two regimes -- in particular, whether they are
exponential or power law in time -- were not investigated quantitatively.
However, the data presented in this work seem consistent with a slower than
exponential decay of the density profile.

In \cite{tien}, an exponential output concentration profile,
$c(x)\sim{\rm exp}(-\Lambda x)$, is assumed from the outset, where $x$ is the 
downstream distance and $\Lambda$ is the experimentally measured filter 
coefficient. The corresponding experimental data show that $\Lambda$ is 
constant at early times, and then
sharply decreases at later times.  Thus the initial stage of the experiment
is consistent with an exponential profile, but later the profile
decays more slowly.  In \cite{tien}, this is attributed to a ``blocking
effect'' in which previously-deposited particles can block the further
deposition of particles onto nearby available trapping sites.

{\em Probability of Encountering an Open Trap}.  As a last remark,
let us examine the assumption that the probability of encountering an open
trap is proportional to the fraction of open traps in a pore
(Sec.~\ref{1dmulti}).  Suppose instead that one takes into account the
volumetric effect that particles far from the surface of the pore do not have
chance to encounter a trap. Then the fraction of particles in contact with
the inner surface of a tube is proportional to $1/r$, namely, the ratio
between the surface area and the volume of the tube.  This would lead to the
interaction terms involving $T_k$ in Eqs.~(\ref{bubblepdf}) and
(\ref{bubblecap}) being multiplied by another factor of $1/r$.  However, this
modification does not affect the propagation velocity of the front, nor the
power law feature of the tail.  Only the decay exponent changes through the
steps of Eqs.~(\ref{tail})-(\ref{tail3}) with an additional factor of $1/r$.

Our results can also provide practical guidelines for improving the design of
a filter in two aspects, namely, the breakthrough time and the amount of
filter material used before the breakthrough.  A longer breakthrough time can
be achieved by having a smaller filter grain size, a lower input
concentration, or a larger pore capacity. While these trends may seem
intuitively clear, we can quantitatively estimate the increase in the
breakthrough time through the expression for the propagation velocity,
Eq.~(\ref{bubblev}).  When breakthrough occurs, the amount of unused filter
material is determined by the shape of the tail in the density profile of the
defenders.  According to Eq.~(\ref{tail4}), the amplitude of this tail is
proportional to $\gamma^{-\lambda}$.  From this, we can quantitatively
estimate the amount of filter material left unused at the breakthrough time
as a function of the reaction strength.

\bigskip {\bf Acknowledgments} We thank Dr.\ Jysoo Lee for helpful
discussions about flow field calculations.  We are also grateful to grants
ARO DAAD19-99-1-0173 and NSF DMR9978902 for financial support.

\begin{appendix}
\section{Solution of the hydrodynamic dispersion equation}\label{appdx}

We solve the master equations, Eq.~(\ref{bubbleme}), by the Laplace transform
method.  Define $\tilde p_k(s)\equiv \int_0^\infty dt\, p_k(t)e^{-st}$, and
take Laplace transform of Eq.~(\ref{bubbleme}) to find
\begin{equation}
\label{laplace}
s\tilde p_0={\rho\phi\over s}-{\tilde p_0\over\tau_0},\quad
s\tilde p_k={\tilde p_{k-1}\over\tau_{k-1}}
       -{\tilde p_k\over\tau_k}.
\end{equation}
Rearranging yields
\begin{equation}
\label{laplace0}
{\tilde p_0\over\tau_0}(1+\tau_0s)={\rho\phi\over s},\quad
{\tilde p_k\over\tau_k}(1+\tau_ks)={\tilde p_{k-1}\over\tau_{k-1}}.
\end{equation}
We multiply the above equations for indices 0, 1, $2,\,\cdots\,\;k$
together and rearrange terms yet again to obtain
\begin{equation}
\label{laplace1}
{\tilde p_k\over\tau_k}\prod_{m=0}^k(1+\tau_ms)={\rho\phi\over s}.
\end{equation}
Before taking averages, we use 
$\tau_k={1\over s}\left[1-1/(1+\tau_ks)\right](1+\tau_ks)$ on the left hand 
side of Eq.~(\ref{laplace1}) to get
\begin{equation}
\label{laplace2}
\tilde p_k\prod_{m=0}^k(1+\tau_ms)={\rho\phi\over s^2}
   \left(1-{1\over 1+\tau_ks}\right)(1+\tau_ks).
\end{equation}
Averaging over the residence time distribution $R(\tau)$ defined by
Eq.~(\ref{ftau}), we obtain
\begin{equation}
\label{pntilde}
\langle{\tilde p}_k\rangle=
   {\rho\phi\over s^2}(1-B)B^k.\qquad B\equiv\left\langle {1\over 1+\tau s}\right\rangle,
\end{equation}
where $\langle\cdot\rangle$ denotes an average over $R(\tau)$.  Note that $B$
does not depend on $m$ because all the $\tau_m$'s are independent and
identically distributed in the large $w$ limit.  For the long-time limit, we
need the small-$s$ behavior of $B$.  Accordingly, we expand $B= 1-
s\langle{\tau}\rangle+ s^2\langle{\tau^2}\rangle\ldots$ in terms
of the moments $\langle{\tau^m}\rangle$.

The profile of the carrier fluid (dashed line in
Fig.~\ref{fig1}) is monotonically decreasing near the front, thus can be
characterized by its first derivative, which gives a 
bell shaped distribution centered at the front. We divide the derivative by 
the total sum of derivatives 
$\sum_m (\langle p_{m}\rangle-\langle p_{m-1}\rangle)=-\langle p_0\rangle$ 
to obtain the normalized probability distribution of the front
\begin{equation}
\label{deriv}
{\cal P}(k)= {1\over\langle p_0\rangle} [\langle p_{k-1}\rangle-\langle p_k\rangle].
\end{equation}

The average position of the front is, using Eq.~(\ref{pntilde}), and the
steady state solution of Eq.~(\ref{bubbleme}), $\langle
p_0\rangle=\rho\phi\langle\tau\rangle$,
\begin{eqnarray}
\label{hydrox}
\bar{\langle{k}\rangle}&=&{1\over\langle p_0\rangle}\sum_k k 
  [\langle p_{k-1}\rangle-\langle p_k\rangle]\nonumber\\
&=&{1\over\langle p_0\rangle} {\cal L}^{-1} \bigl(\sum_k k
    [\langle\tilde p_{k-1}\rangle-\langle\tilde p_k\rangle]\bigr)\nonumber\\
&=&{1\over\langle\tau\rangle}{\cal L}^{-1}\Bigl({1-B\over s^2}
    \sum_k k[B^{k-1}-B^k]\Bigr)\nonumber\\
&\simeq&{t\over \langle{\tau}\rangle},
\end{eqnarray}
where the over-bar means averaging over ${\cal P}(k)$, and 
${\cal L}^{-1}$ is the inverse Laplace transform. We use the identity 
${\cal L}^{-1}(s^{-\beta})=t^{\beta-1}/\Gamma(\beta)$ for 
the last step. We assume
sufficiently long chain of bubbles in the summations above to prevent finite
length effect.  From above, we find the propagation velocity as
$v=\langle\tau\rangle^{-1}$ (Eq.~(\ref{hydrov})).

Similarly, $\bar{\langle{k^2}\rangle}=\langle p_0\rangle^{-1}\sum_kk^2
[\langle p_{k-1}\rangle-\langle p_k\rangle]$, and the width of the front is
\begin{equation}
\label{hydrow1}
[\bar{\langle{k^2}\rangle}-\bar{\langle{k}\rangle}^2]^{1\over2}
  \simeq\left\{{t\over\langle{\tau}\rangle}\left[
  2\Gamma(1+{\mu\over2})\Gamma(3-{\mu\over2})-1\right]\right\}^{1\over2},
\end{equation}
which gives Eq.~(\ref{hydrow}) 
\end{appendix}
\end{multicols}

\begin{thebibliography}{99}
\bibitem{revs} C. Tien and A. C. Payatakes, AIChE J. {\bf 25}, 737 (1979).

\bibitem{probstein} R. F. Probstein, {\it Physicochemical Hydrodynamics},
2nd ed. (John Wiley \& Sons, 1994).

\bibitem{sahimi} A. O. Imdakm and M. Sahimi, Phys.\ Rev.\ A {\bf 36},
  5304 (1987).

\bibitem{luhrmann} L. L\"{u}hrmann, U. Noseck, and C. Tix,
Water Res.\ Res.\ {\bf 34}, 421 (1998).

\bibitem{bloomfield} L. Bloomfield, Scientific American, p152
(December, 1999).

\bibitem{putnam} D. Putnam and M. Burns, Chem.\ Eng.\ Sci.\ {\bf 52},
  93 (1997).

\bibitem{brenner} H. Brenner and D. A. Edwards, 
   {\it Macrotransport Processes} (Butterworth-Heinemann, 1993).

\bibitem{rajagopalan} R. Rajagopalan and C. Tien, AIChE J. {\bf 22},
  523 (1976).

\bibitem{jysoo} J. Lee and J. Koplik,  Phys.\ Rev.\ E {\bf 54}, 4011 (1996).

\bibitem{ghidaglia} C. Ghidaglia, L. de Arcangelis, J. Hinch, and E. Guazzelli,
Phys.\ Rev.\ E {\bf 53}, R3028 (1996).

\bibitem{datta} S. Datta and S. Redner, Phys.\ Rev.\ E {\bf 58}, R1203 (1998);
  S. Datta and S. Redner, Int.\ J. Mod.\ Phys.\ C {\bf 9} 1535 (1998).

\bibitem{imdakm} A. O. Imdakm and M. Sahimi, Chem.\ Eng.\ Sci.\ {\bf 46}, 
  1977 (1991).

\bibitem{rege1} S. Rege and H. Fogler, Chem.\ Eng.\ Sci.\ {\bf42}, 
  1553 (1987).


\bibitem{chiang} H. Chiang and C. Tien, AIChE J. {\bf31}, 1349 (1985).

\bibitem{levich} V. G. Levich, {\it Physicochemical Hydrodynamics}
(Prentice-Hall, 1962).

\bibitem{bubbles} H. E. Daniels, Proc.\ R. Soc.\ London, Ser. A {\bf 183}, 
404 (1945); D. G. Harlow and S. L. Phoenix, J. Comput.\ Mater.\ {\bf 12},
195 (1978);
S. L. Phoenix and R. L. Smith, J. Appl.\ Mech.\ {\bf 103},
75 (1981); D. Sornette and S. Redner, J. Phys.\ A {\bf 22}, L619 (1989);
P. L. Leath and P. M. Duxbury, Phys.\ Rev.\ B {\bf 49}, 14905 (1994).

\bibitem{kahng} B. Kahng, G. G. Batrouni, and S. Redner, J. Phys.\ A
{\bf 20} L827 (1987).

\bibitem{thomas} G. H. Thomas, G. R. Countryman, and I. Fatt, Soc.\ Pet.\ 
Eng.\ J. {\bf 3}, 189 (1963); I. Chatzis and F. A. L. Dullien, Int.\ Chem.\ Eng.\
{\bf 25}, 47 (1985).

\bibitem{redner} J. Koplik, S. Redner, and E. J. Hinch, Phys.\ Rev.\ E
{\bf 50}, 4650 (1994)

\bibitem{pdf} L. Niemeyer, L. Pietronero, and H. J. Wiesmann, 
Phys.\ Rev.\ Lett.\ {\bf 52}, 1033 (1984); S. Havlin, M. Dishon, J. E. Kiefer,
and G. H. Weiss, Phys.\ Rev.\ Lett.\ {\bf 53}, 407 (1984).

\bibitem{thesis} W. Hwang, PhD thesis (Boston University, 2000).

\bibitem{numrecipe} W. H. Press, S. A. Teukolsky, W. T. Vetterling, and
  B. P. Flannery, {\it Numerical Recipes in C}, 2nd ed. (Cambridge, 1999).

\bibitem{tien} R. Vaidyanathan and C. Tien, Chem.\ Eng.\ Sci.\ {\bf 46},
967 (1991).


\end{thebibliography}
\end{document}